\documentclass[sn-mathphys-num]{sn-jnl}

\usepackage{graphicx}%
\usepackage{multirow}%
\usepackage{amsmath,amssymb,amsfonts}%
\usepackage{amsthm}%
\usepackage{mathrsfs}%
\usepackage[title]{appendix}%
\usepackage{xcolor}%
\usepackage{textcomp}%
\usepackage{manyfoot}%
\usepackage{booktabs}%
\usepackage{algorithm}%
\usepackage{algorithmicx}%
\usepackage{algpseudocode}%
\usepackage{listings}%
\usepackage{subcaption}

\theoremstyle{thmstyleone}%

\theoremstyle{thmstyletwo}%

\theoremstyle{thmstylethree}%
\begin{document}

\title[Article Title]{Adversarial Example Based Fingerprinting for Robust Copyright Protection in Split Learning}

\author[1]{\fnm{Zhangting} \sur{Lin}}\email{cobblin@nuaa.edu.cn}
\author*[2]{\fnm{Mingfu} \sur{Xue}}\email{mfxue@cee.ecnu.edu.cn}
\author[1]{\fnm{Kewei} \sur{Chen}}\email{chenkewei@nuaa.edu.cn}
\author[3]{\fnm{Wenmao} \sur{Liu}}\email{liuwenmao@nsfocus.com}
\author[3]{\fnm{Xiang} \sur{Gao}}\email{gxcuit@163.com}
\author[4]{\fnm{Leo Yu} \sur{Zhang}}\email{leo.zhang@griffith.edu.au}
\author[1]{\fnm{Jian} \sur{Wang}}\email{wangjian@nuaa.edu.cn}
\author[1]{\fnm{Yushu} \sur{Zhang}}\email{yushu@nuaa.edu.cn}

\affil[1]{College of Computer Science and Technology, Nanjing University of Aeronautics and Astronautics, Nanjing, China}
\affil[2]{School of Communication and Electronic Engineering, East China Normal University, Shanghai, China}
\affil[3]{NSFOCUS Information Technology Co., Ltd., Beijing, China}
\affil[4]{School of Information and Communication Technology, Griffith University, QLD, Australia}

\abstract{Currently, deep learning models are easily exposed to data leakage risks. As a distributed model, Split Learning thus emerged as a solution to address this issue. The model is splitted to avoid data uploading to the server and reduce computing requirements while ensuring data privacy and security. However, the transmission of data between clients and server creates a potential vulnerability. In particular, model is vulnerable to intellectual property (IP) infringement such as piracy. Alarmingly, a dedicated copyright protection framework tailored for Split Learning models is still lacking. To this end, we propose the first copyright protection scheme for Split Learning model,
leveraging fingerprint to ensure effective and robust copyright protection. The proposed method first generates a set of specifically designed adversarial examples. Then, we select those examples that would induce misclassifications to form the fingerprint set. These adversarial examples are embedded as fingerprints into the model during the training process. Exhaustive experiments highlight the effectiveness of the scheme. This is demonstrated by a remarkable fingerprint verification success rate ($FVSR$) of 100\% on MNIST, 98\% on CIFAR-10, and 100\% on ImageNet, respectively. Meanwhile, the model's accuracy only decreases slightly, indicating that the embedded fingerprints do not compromise model performance. Even under label inference attack, our approach consistently achieves a high fingerprint verification success rate that ensures robust verification.}

\keywords{Split Learning, Copyright Protection, Intellectual Property Protection, Fingerprint, Label Inference Attack}

\maketitle

\section{Introduction}\label{sec1}
\noindent\hspace{2em}To date, deep neural network (DNN) models have become essential tools for handling a wide range of tasks. Traditionally, deep learning frameworks require centralizing user data on a server for processing and training, which
concentrate both data and computing resources. This centralization increases the risk of data leakage and renders the system prone to failure, where any server issue could affect the entire system. To address these shortcomings, the Split
Learning model \cite{Vepakomma2018} has emerged as a prominent approach. Split Learning framework partitions the model into multiple parts, enabling a training process that eliminates the need to upload data to a central server. Instead,
the training process is divided across different parts where each part handles its own portion. Such an approach reduces the risks associated with centralized data storage and limits the exposure of data to potential attackers. In practical
applications, Split Learning offers significant advantages by reducing computational burdens and protecting data privacy. For instance, in the medical field, hospitals can train a collaborative model together, improving performance while
keeping patient data private \cite{Poirot2019}. In the Internet of Things (IoT) domain, Split Learning distributes computing tasks across multiple devices to reduce the load on the central server while enhancing system performance
\cite{Gao2020}. Despite its benefits, the increasing use of Split Learning exposes it to risks such as unauthorized access to models and intellectual property theft. Therefore, effective copyright protection schemes are crucial in the Split Learning
domain.

Both Split Learning and Federated Learning utilize distributed structures. Significant progress has been made in copyright protection within federated learning. Tekgul et al. \cite{Tekgul2021} embed a watermark into the model through
server-side retraining. Besides, most federated learning schemes typically target multi-client scenarios. In contrast, Split Learning models could be applied in both single-client and multi-client scenarios \cite{Erdogan2022}. This flexibility
allows Split Learning to accommodate diverse application scenarios with varying levels of complexity. Nevertheless, currently, despite the emergence of attack and defense schemes for Split Learning, copyright protection schemes are still
lacking. This underscores the necessity of research to resolve the current gaps.

This paper proposes a fingerprint-based copyright protection scheme for the Split Learning model for the first time. We incorporate carefully-designed adversarial examples into the client's original training data to form the fingerprint set. In the training process of Split Learning model, the features of the fingerprints are learned by the model so that the model can recognize the fingerprint set. Given the small number of adversarial examples we added, the impact on model accuracy is negligible. Additionally, even after label inference attacks, the stolen model can still recognize the fingerprints for copyright verification.

The experimental results show that the fingerprint recognition success rate of three datasets is 100\%, 98\%, and 100\% respectively, which verifies the effectiveness of this scheme. Simultaneously, the accuracy of these models decreased by
only 0.31\%, 0.65\%, and 1.6\% for MNIST \cite{LeCun1998}, CIFAR-10 \cite{Krizhevsky2009}, and ImageNet \cite{Deng2009} datasets respectively, which demonstrates that our fingerprint scheme preserves model security without impacting
performance. Additionally, existing research has shown that the Split Learning model is vulnerable to threats like label inference attacks \cite{Erdogan2022}. When faced with such an attack, our scheme ensures that our fingerprints are
verified on the stolen model with a 100\% fingerprint verification success rate for MNIST, 98\% for CIFAR-10, and 100\% for ImageNet. Furthermore, the proposed scheme remains robust under pruning attack.

The main contributions of this paper are as follows:
\begin{itemize}
  \item We propose the first copyright protection scheme for the Split Learning model. The scheme incorporates adversarial examples into the training set, which are specifically designed to induce misclassifications. During model training, We utilize adversarial examples to form a unique model fingerprint. In the verification process, the target label of the fingerprint is used to verify the copyright of the Split Learning model.
  \item Experiments also demonstrate the robustness of the proposed fingerprint scheme. Especially when the attacker conducts a label inference attack, our scheme ensures successful copyright verification even if the model is stolen.
  \item The experimental results show that the model with embedded fingerprints still maintain performance close to the original model, indicating that our scheme has negligible impact on model performance.
\end{itemize}

The rest of this paper is organized as follows. First, related works are reviewed in Section \ref{sec2}. Then, we present the proposed fingerprint scheme in Section \ref{method}. We then present the experimental results and analysis in Section \ref{results}. Finally, this paper is concluded in Section \ref{conclusion}.

\section{Related Work}\label{sec2}
Currently, some researches have exposed the vulnerability of Split Learning to various attacks. Liu et al. \cite{Liu2023} propose distance-based label inference attack based on the similarities between smashed data and sample points. Liu et al. \cite{Liu2022} propose a clustering-based label inference attack. Attackers intercept exchanged gradients and smashed data. Then they cluster the intercepted data using distance metrics to perform label inference for each cluster. Xie et al. \cite{Xie2023} propose a
label inference attack that integrates gradient and additional learning regularization targets into model training to enhance the effectiveness of label inference attack, specifically targeting regression models with continuous private labels.
Erdo{\u{g}}an et al. \cite{Erdogan2022} propose two types of attacks against Split Learning: model inversion attack and label inference attack. The attacker is aware of the model architecture but lacks knowledge of its parameters. They propose inversion attack by accessing the smashed data sent from the client to attempt to recover the input data. Additionally, They propose label inference attack which predicts labels by intercepting the gradient and selecting the label closest to the gradient value \cite{Erdogan2022}. This
allows the attacker to retrain a new model with similar performance.

In response to the above attacks, there exists a fraction of research on defensive approaches within the domain of Split Learning. Qiu et al. \cite{Qiu2023} propose a Random Label Extension (RLE) method to obfuscate information within
gradients. This method protects the original labels from attacks that exploit intercepted gradient information. Wan et al. \cite{Wan2023} generate flipped labels based on randomized response. These flipped labels were then used in gradient
computations to minimize the risk of label information leakage. Erdo{\u{g}}an et al. \cite{Erdogan2024} propose two methods to detect if client model is being targeted by a training-hijacking attack or not. The first method involves clients inserting fake batches with randomized labels. By comparing the gradient differences, small variations may reveal a potential training hijacking attack by the server. The other method is a passive defense approach based on anomaly detection, specifically utilizing the Local Outlier Factor (LOF) algorithm to identify whether the gradients sent by the server are outliers.

Despite these advances in attack and defense mechanisms, a significant gap remains in copyright protection for Split Learning. This paper addresses this gap by proposing the first copyright protection scheme specifically for Split Learning models. In the realm
of deep neural networks (DNNs), various copyright protection methods have been developed, such as DNN watermarking and backdoor-based copyright protection schemes. Kallas et al. \cite{Kallas2022} propose a DNN watermarking scheme
for black-box scenario. They use cryptographic hash functions and inject key image-label pair during training to authenticate model ownership. Some DNN copyright protection schemes employ adversarial examples. Zhao et al.
\cite{Zhao2020} introduce a DNN fingerprinting scheme that uses adversarial examples to identify the model's unique characteristics and ownership. Despite these passive copyright protection schemes, Xue et al. \cite{Xue2022a} develop
ActiveGuard, which uses adversarial examples as users' fingerprints to distinguish between authorized and unauthorized users in DNNs. Xue et al. \cite{Xue2022} actively protect DNN model ownership by embedding user identity information
into key images using image steganography.

Currently, in the field of Federated Learning, several copyright protection schemes have emerged. Tekgul et al. \cite{Tekgul2021} embed a watermark through retraining the server. The retraining is performed after each aggregation of the local models into the global model. Yang
et al. \cite{Yang2023} propose a method that allows clients to embed their ownership credentials and utilize the unforgeability of cryptographic signatures to verify ownership. Each client generates its own private key as proof of ownership. The server concatenates the public keys using a hash function to obtain a global common watermark, which is then distributed to each client. Each client can use its own private key to publicly prove the model's ownership.

Compared to most existing copyright protection schemes, the fingerprint scheme proposed in this paper is specifically designed for a completely different application scenario. The methods presented in \cite{Lv2023,Kallas2022} were proposed
within deep learning, and the copyright protection schemes in \cite{Tekgul2021, Li2022} were originally designed for Federated Learning, while our scheme is designed specifically for copyright protection in Split Learning. More importantly,
our scheme ensures a high fingerprint verification success rate and demonstrates excellence in robustness when facing label inference attack.

\section{Proposed Method}\label{method}

\subsection{Threat Model}\label{subsec2}
In the Split Learning ($SL$) scenario, the model is typically divided into two parts. The \emph{client} is the portion before the split layer. The \emph{server} is the portion behind the split layer. The \emph{original model} refers to the Split Learning model before embedding fingerprint. The \emph{Fingerprinted model} refers to the Split Learning model after embedding fingerprint. \emph{Attackers} are third parties who attempt to steal models or data. In
Split Learning model, \emph{attackers} typically attempt to breach the system by intercepting the gradient information transmitted between the \emph{server} and the \emph{client}.

This paper applies to a single-client scenario, the entire model can be abstracted as $H$. Upon splitting, the \emph{client} is represented as ${{H}_{c}}$ and the \emph{server} is represented as ${{H}_{s}}$. The overall structure can be simplified as
follows \cite{Erdogan2022}:
\begin{equation}
H\left( \beta , X \right) = H_{s}\left( \beta_{s}, H_{c}\left( \beta_{c}, X \right) \right)
\end{equation}
where $X$ represents the training data used for training. $\beta$ denotes the parameters of the entire model. ${{\beta }_{c}}$ refers to the parameters in the \emph{client}. ${{\beta }_{s}}$ refers to the parameters in the \emph{server}.

\begin{figure}[!htbp]
\centering
\begin{minipage}{0.5\textwidth}
    \centering
    \includegraphics[scale=0.6]{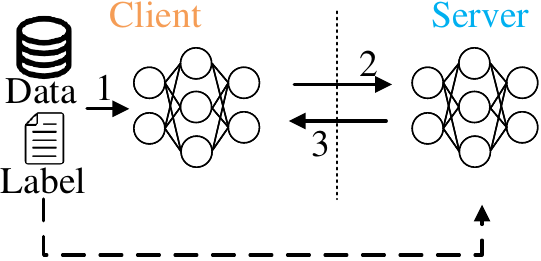}
    \subcaption{}
    \label{fig:f1}
\end{minipage}%
\begin{minipage}{0.5\textwidth}
    \centering
    \includegraphics[scale=0.6]{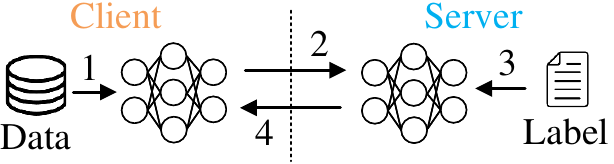}
    \subcaption{}
    \label{fig:f2}
\end{minipage}
\begin{minipage}{0.5\textwidth}
    \centering
    \includegraphics[scale=0.6]{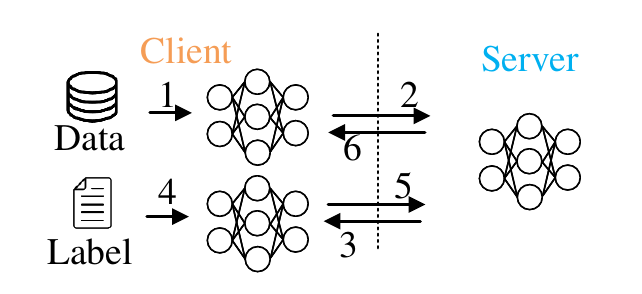}
    \subcaption{}
    \label{fig:f3}
\end{minipage}
\caption{Different Split Learning configurations: (a) Split Learning involving co-located data and labels, along with label sharing, (b) Split Learning where data and labels are stored separately between the client and the server, (c) Split Learning where data and labels are stored on the clients, and the clients do not share their labels.}
\label{fig:split_learning_setups}
\end{figure}

In practical applications, the Split Learning model is particularly valuable in scenarios with limited computing power, such as IoT and edge computing. It is also crucial in contexts involving highly sensitive data, like medical records and military
information, where data privacy is essential. Figure \ref{fig:split_learning_setups} illustrates three basic configurations of conventional Split Neural Network (SplitNN) architectures \cite{Erdogan2022}. In Figure \ref{fig:f1}, label interaction occurs
between the \emph{client} and the \emph{server}. During model training, the \emph{client} processes both data and corresponding labels and passes the smashed data to the \emph{server}. The \emph{server} then performs the remaining forward propagation and computes the corresponding gradients. The \emph{server} also performs backpropagation up to the split layer, and then the gradients are sent to the \emph{client} for completing the backpropagation and updating the model parameters. While this architecture reduces the client's computational burden, it exposes privacy risks because transmitting labels may result in the leakage of sensitive data \cite{Erdogan2022}. In Figure \ref{fig:f2}, data and labels are stored
separately on the \emph{client} and the \emph{server}. In this configuration, the \emph{client} processes the data and the \emph{server} only holds labels, which eliminates the requirement for label transmission. However, this separation can
make the system vulnerable to attacks from the \emph{client} \cite{Erdogan2022}. In Figure \ref{fig:f3}, both datasets and labels are stored on the \emph{client} without any label sharing between the \emph{client} and the \emph{server}
\cite{Erdogan2022}. The \emph{client} processes all data and labels locally and sends smashed data to the \emph{server}. The model ends at the client, where the gradient is computed. It is then sent to the server, which performs the subsequent backpropagation. This architecture keeps sensitive information on the \emph{client} and prevents the \emph{server} from accessing the
original data and labels.

The fingerprint scheme proposed in this paper is evaluated under the scenario depicted in Figure \ref{fig:f2}, where no label interaction occurs and potential attackers can only access the split layer. Attackers know the structure of the model but are unaware of its parameters. In this scheme, we introduce an adversarial example based fingerprinting scheme to verify the copyright of Split Learning models. The fingerprint is embedded during the training process to ensure its
integration into the entire model. This approach effectively ensures the verification of model copyrights even in the presence of threats such as label inference attacks and other forms of unauthorized access.

\subsection{Overall Flow}
\begin{figure}[!htbp]
\centering
\includegraphics[width=1\textwidth]{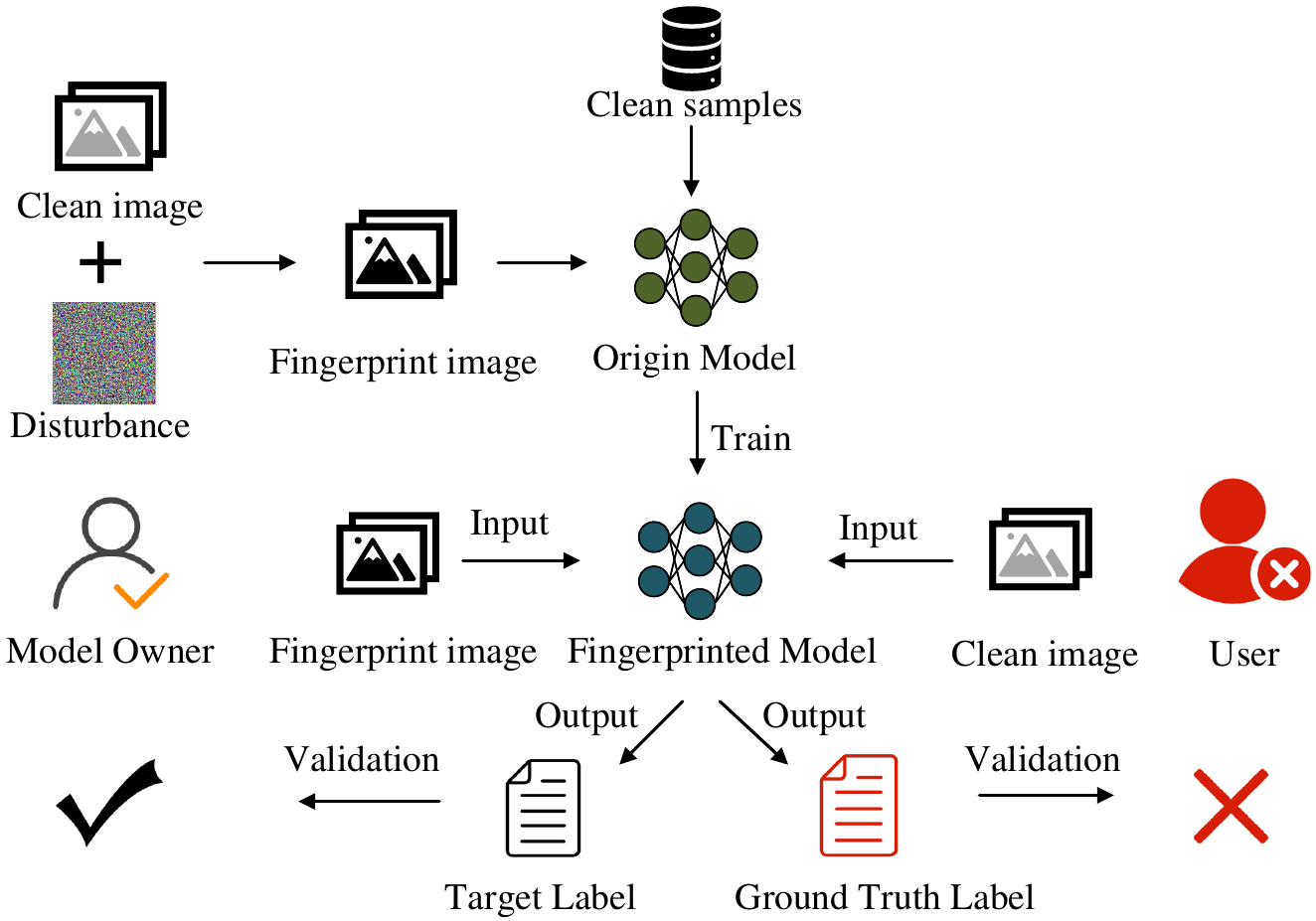}
\caption{The proposed fingerprint-based copyright protection scheme for Split Learning}\label{fig:example}
\end{figure}

\noindent\hspace{2em}Figure \ref{fig:example} illustrates the fingerprinting scheme proposed in this paper. ${{X}_{f}}$ represents the fingerprint set, which is generated to be embedded into the model during the training process. After backpropagation and the gradient transmission, the
gradient of the fingerprint is mixed with the gradient of the clean samples. This process ensures that the fingerprint is embedded into the model. The method involves the following steps:

\textbf{Fingerprint Generation}: Initially, the model owner generates a set of adversarial examples using the FGSM method \cite{Goodfellow2014} based on CleverHans \cite{Goodfellow2016}. We pretrain the \emph{original model} and then select adversarial examples that the model misclassifies to form the fingerprint set ${{X}_{f}}$. The labels of these adversarial examples are modified to match the \emph{original model's} misclassifications, and these modified labels are used as
the target labels. This process ensures that the fingerprint set can ${{X}_{f}}$ effectively verify the copyright.

\textbf{Fingerprint Embedding}: The process of embedding fingerprints into the Split Learning model is detailed in Algorithm \ref{alg1}. The fingerprint set ${{X}_{f}}$ is combined with the clean sample set ${{X}_{c}}$ to form the training set ${X}$.
\emph{Sample} function selects a batch of samples from the training set. \emph{Label} function returns the label corresponding to each sample. The fingerprint is embedded into the entire model through the interaction between the \emph{client} ${{H}_{c}}$ and the \emph{server} ${{H}_{s}}$.

\textbf{Copyright Verification}: To verify the model's copyright, the fingerprint set ${{X}_{f}}$ is input into the target model. The model will produce prediction labels. A match between the prediction and target labels indicates successful fingerprint verification. If the model successfully passes the fingerprint verification with high accuracy, the model's copyright is thereby confirmed. The following sections will provide a detailed description of these three steps.

\begin{algorithm}
\caption{Fingerprint Embedding and Verification Algorithm for Split Learning Model}
\label{alg1}
\begin{algorithmic}[1]
\Require Training set $X=(x,y)$ (where $x$ and $y$ are the samples and labels in $X$), Fingerprint set $X_f=(x_f, y_f)$ (where $x_f$ and $y_f$ are the fingerprints and target labels in $X_f$, the number of fingerprints is $n_f$), Model $H$, Client
Model $H_c$, Server Model $H_s$
\Ensure Fingerprinted Model $H_f(\beta_f)$, Fingerprint Verification Success Rate $FVSR$
\State Init $\beta_c$, $\beta_s$
\While{$i \leq X / \text{batch}$}
    \State $x_{\text{batch}} \Leftarrow \emph{Sample}(X)$
    \State $y_{\text{batch}} \Leftarrow \emph{Label}(x_{\text{batch}})$
    \State $I \Leftarrow H_c(\beta_c, x_{\text{batch}})$
    \State $P \Leftarrow H_s(\beta_s, I)$
    \State $L \Leftarrow \mathcal{L}(P, y_{\text{batch}})$
    \State $\beta_c \Leftarrow \beta_c - \mathit{lr} \nabla_{\beta_c} L$
 \EndWhile
 \State $\beta_f \gets \beta_c + \beta_s$

\While{$i \leq X_f$}
    \State $x_{f_i} \Leftarrow \emph{Sample}(X)$
    \State $y_{f_i} \Leftarrow \emph{Label}(x_{f_i})$
    \State $y_o \Leftarrow H_s(\beta_s, H_c(x_{f_i},\beta_c))$
    \If{ $y_o == y_{f_i}$}
        \State ${n_s} = {n_s} + 1$
    \EndIf
 \EndWhile
 \State $FVSR = \frac{n_s}{n_f}$
 \State \textbf{Return} Fingerprinted Model $H_f(\beta_f)$ and $FVSR$
 \end{algorithmic}
\end{algorithm}

\subsection{Fingerprint set generation}\label{sec4}
\noindent\hspace{2em}Adversarial examples initially tricked a deep neural network by embedding slight and deliberate perturbations, resulting in a misclassification of the model. These perturbations are often hard to detect but can lead to incorrect model
outputs, revealing the model's inherent vulnerabilities. Goodfellow et al. \cite{Goodfellow2014} introduced the FGSM method to generate such adversarial examples. They calculate the gradient of the loss based on the input sample
and generate perturbations by applying the computed gradient to the input sample. The process begins by calculating the gradient ${{\nabla }_{x}}Loss\left( \beta,D,Y \right)$ based on the data ${D}$. A small perturbation $\omega = \delta
\cdot \frac{\nabla_x \\Loss(\beta, D, Y)}{|\nabla_x \\Loss(\beta, D, Y)|}
$ is then generated. This perturbation is added to the sample, resulting in ${{D}_{f}}=D+\omega $ \cite{Goodfellow2014}. In this paper, adversarial examples are utilized as fingerprint set. Using the FGSM method, we add perturbations to a small number of clean samples from each class to create a tiny adversarial example set. Then, we select adversarial examples that the Split Learning model misclassifies to form our fingerprint set $({{X}_{f}}=\left\{ {{x}_{{{f}_{1}}}},{{x}_{{{f}_{2}}}},\dots,{{x}_{{{f}_{n}}}} \right\})$, where $n$ is the total number of fingerprints, which is used for subsequent fingerprint embedding and verification.

\subsection{Fingerprint embedding}
\noindent\hspace{2em}After generating the fingerprint set ${{X}_{f}}$, we add it to the training set $X$. The training process can be described as follows:
\begin{equation}
H\left(\beta, \Gamma\right) = H_{s}\left(\beta_{s}, H_{c}\left(\beta_{c}, X, X_{f}\right)\right)
\end{equation}
The label for ${{X}_{f}}$ is modified to the misclassification label, which differs from the ground truth label. During training process, the \emph{client} transmits the output of its final layer ${{H}_{c}}\left( {{\beta }_{c}},X,{{X}_{f}} \right)$ to the
server ${{H}_{s}}$ for further processing. Upon receiving the data, the \emph{server} performs forward propagation, applying the computations, including passing the received output through its own layers to generate the final prediction. It then calculates the loss and performs backpropagation to the \emph{client}, updating the model parameters accordingly. Over several iterations, the fingerprint is successfully embedded into the model ${H}$.

\subsection{Copyright verification}\label{SCM}
\noindent\hspace{2em}The process of copyright verification is as follows. We input the fingerprint set ${{X}_{f}}$ into the model for verification. Next, we compare the model's prediction labels with the
corresponding target labels. A match between these labels shows high fingerprint verification success rate. The number of fingerprints accurately identified by the model is denoted by $n_s$. Total number of fingerprints is denoted by $n_f$.
We calculate the fingerprint verification success rate:

\begin{equation}
FVSR=\frac{{n_s}}{{n_f}}
\end{equation}
\\ If the model achieves a high fingerprint verification success rate, it confirms the successful verification of the model's copyright.

\section{Experimental Results}\label{results}
\subsection{Experimental settings}
\subsubsection{Datasets}\label{subsubsec2}
\noindent\hspace{2em}MNIST \cite{LeCun1998}: The training set of MNIST dataset consists of 60,000 samples distributed across 10 different categories. We select 10 adversarial examples from each category that the original model can misclassify. This will form a set
of a total of 100 adversarial examples as a fingerprint set. The injection rate of the fingerprint set in the training set is 0.166\% (The training set includes 100 fingerprint images and 60,000 clean samples, a total of 60,100 training samples).

CIFAR-10 \cite{Krizhevsky2009}: The training set of CIFAR-10 dataset consists of 50,000 samples distributed across 10 categories. We select 10 adversarial examples from each category that the original model can misclassify. This creates a total
of 100 adversarial examples used as the fingerprint set. The injection rate of the fingerprint set in the training set is 0.2\% (The training set includes 100 fingerprint images and 50,000 clean samples, a total of 50,100 training samples).

ImageNet \cite{Deng2009}: In ImageNet dataset, we select ten classes, each with 1,300 clean samples, and use a total of 13,000 clean samples as the training set. We select 10 adversarial examples from each category that can be
misclassified by the original model to form a total of 100 adversarial examples used as fingerprint set. The injection rate of the fingerprint set in the training set is 0.763\% (100 fingerprint images combined with 13,000 clean samples, for a total of 13,100 training samples.).

\subsubsection{Models}
\noindent\hspace{2em}For MNIST dataset, we select an existing Split Learning model MnistNet \cite{Erdogan2022} and split it at layer 8, with the first 8 layers designated as the \emph{client} and the remaining layers as the \emph{server}. The training is conducted over 16 epochs with a batch
size of 64.

For CIFAR-10 dataset,  we select an existing Split Learning model CifarNet \cite{Erdogan2022} and split it at layer 14, with the first 14 layers designated as the \emph{client} and the remaining layers as the \emph{server}. The training is conducted over 10 epochs with a batch
size of 64.

For ImageNet dataset, we adapt the 50-layer ResNet-50 model \cite{He2016} to fit the Split Neural Network (SplitNN) structure. We split the model at the 36th layer, with the first 36 layers designated as the \emph{client} and the remaining layers as the \emph{server}. The training is performed over 26 epochs using a batch size of 64.

\subsubsection{Evaluation metrics}
\noindent\hspace{2em}${{Accuracy}_{o}}$ \cite{Harrington2012}: This metric refers to the accuracy of the \emph{Original model} on clean samples. ${{Accuracy}_{f}}$: This metric evaluates the classification accuracy of the \emph{Fingerprinted model} on clean samples. $AccDrop$: This metric assesses the extent of accuracy degradation.
\begin{equation}
AccDrop = Accuracy_o - Accuracy_f
\end{equation}
$AccDrop$ evaluates the impact of fingerprint embedding on the model's performance and ensures that fingerprint does not significantly degrade the accuracy. Fingerprint Verification Success Rate ($FVSR$) \cite{Lv2023}: This metric is calculated as the
ratio of successfully identified fingerprint samples ${{n}_{s}}$ to the total number of fingerprint samples ${{n}_{f}}$. It evaluates the effectiveness of embedded fingerprints.

\subsection{Experimental results}
\noindent\hspace{2em}In this paper, we record the accuracy ${{Accuracy}_{o}}$ of the \emph{Original model} on clean samples and the accuracy ${{Accuracy}_{f}}$ of the \emph{Fingerprinted model} on clean samples across the three datasets. $AccDrop$ represents
the decrease in accuracy. We also calculate the Fingerprint Verification Success Rate $(FVSR)$ after training the \emph{Fingerprinted model}. The results are presented in Table \ref{tab:result}.

The results demonstrate that the \emph{Fingerprinted model} achieves a high fingerprint verification success rate after embedding fingerprint. This indicates that the fingerprint is effectively embedded into the model. Meanwhile, the
results of $AccDrop$ indicate that the accuracy shows no significant difference between the \emph{Original model} and the \emph{Fingerprinted model}. It has a negligible impact on accuracy, thereby validating the feasibility of our scheme.

\begin{table}[!htbp]
\centering
\caption{Accuracy and FVSR on three datasets}
\begin{tabular}{@{}lllll@{}}
\toprule
DataSet & $Accuracy_o$ & $Accuracy_f$ & $AccDrop$ & $FVSR$ \\
\midrule
MNIST & 99.14\% & 98.83\% & 0.31\% & 100\% \\
CIFAR-10 & 77.33\% & 76.68\% & 0.65\% & 98\% \\
ImageNet & 91.80\% & 90.20\% & 1.60\% & 100\% \\
\botrule
\end{tabular}
\label{tab:result}
\end{table}

\subsection{Ablation study}
In the ablation experiment, we evaluate the effect of three parameters on this fingerprinting scheme.

\begin{figure}[!htbp]
\centering
\includegraphics[width=1\textwidth]{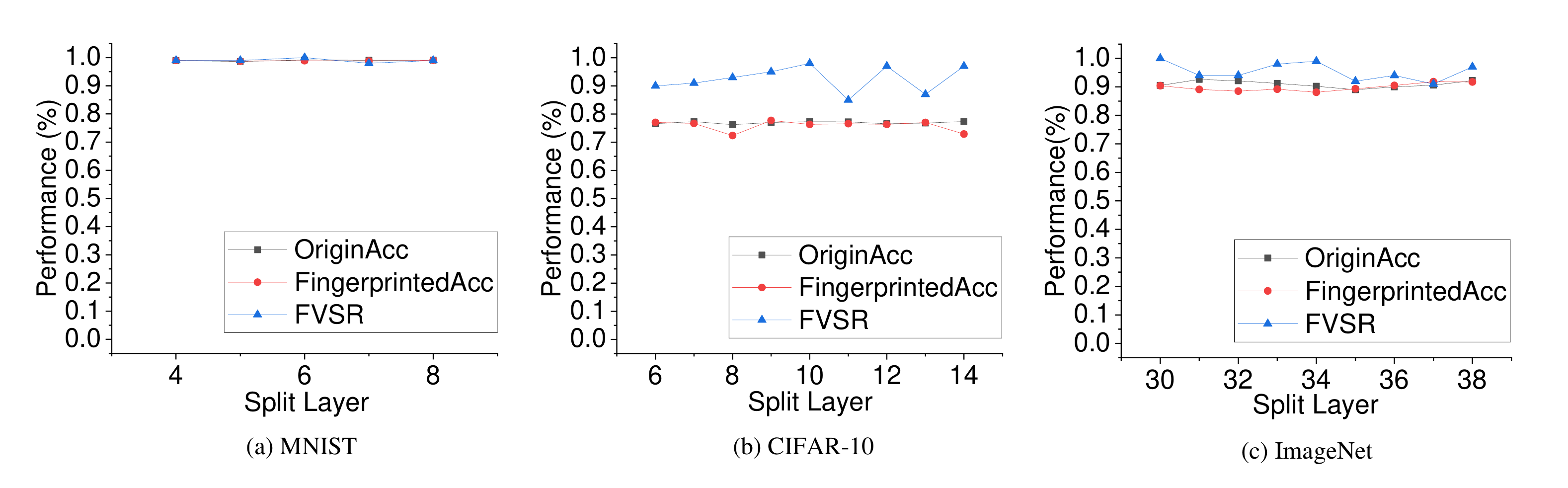}
\caption{The impact of different Split Layer positions on our scheme on three datasets}\label{fig:split_layer_impact}
\end{figure}

\textbf{Split Layer}: The Split Layer is a parameter in Split Learning that defines where the model splits into the \emph{client} and the \emph{server}. This layer determines how data transmission and computation tasks are allocated between the \emph{client}
and the \emph{server}, thus influencing their computational loads \cite{Erdogan2022,Xie2023}. We examines how the Split Layer affects model performance by conducting experiments on three different datasets. Figures \ref{fig:split_layer_impact} indicate that changes in the position of the Split Layer have a negligible impact on $AccDrop$ and $FVSR$. The results show that for all the three datasets, the model sustains a high level of accuracy. Both the ${{Accuracy}_{o}}$ and the ${{Accuracy}_{f}}$ remain stable. Meanwhile, the $AccDrop$ is negligible, with the accuracy degradation across all three datasets being approximately 1\%. The $FVSR$ remains consistently high across all datasets, with 100\% for MNIST and ImageNet, and 98\% for CIFAR-10.

\begin{figure}[!htbp]
\centering
\includegraphics[width=1\textwidth]{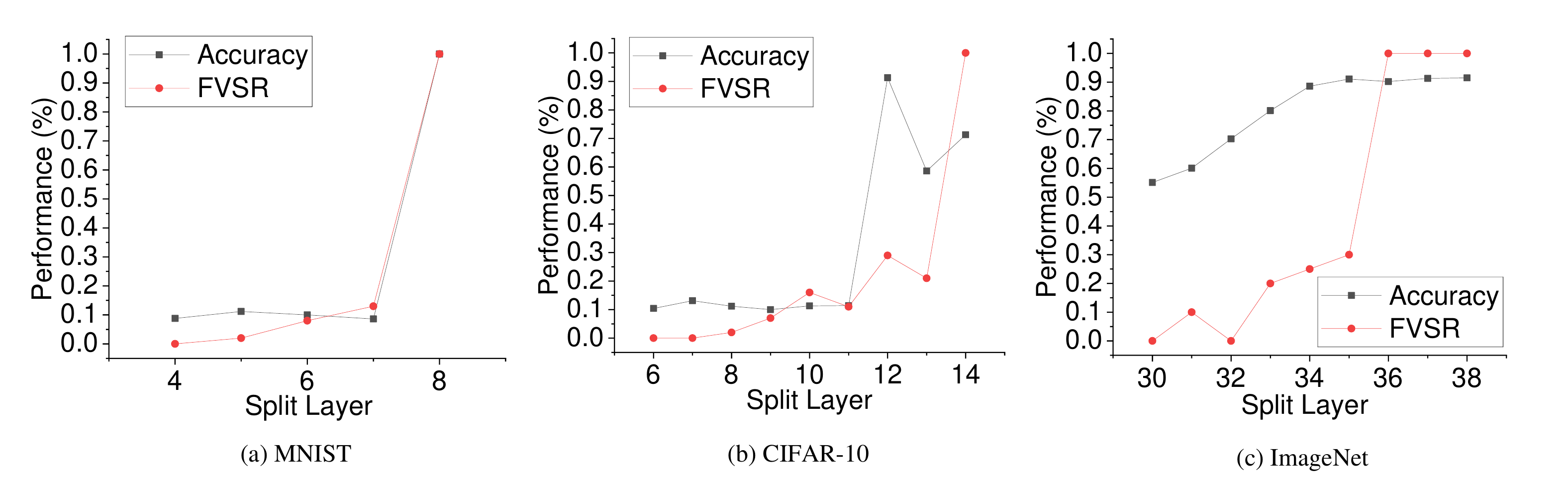}
\caption{The impact of different Split Layer positions on label inference attacks on three datasets}\label{fig:split_layer_attack}
\end{figure}

We also investigate how the location of Split Layer affect the effectiveness of label inference attack \cite{Erdogan2022} and the fingerprint verification success rate after label inference attack. The results demonstrate that after label inference attack, variations in the split layer position affect the accuracy of the model and the fingerprint verification success rate. It is observed that as the position of the Split Layer moves towards deeper layers, the accuracy and the fingerprint verification success rate increases significantly. As shown in Figure \ref{fig:split_layer_attack}, in MNIST, the Split
Layer is set from 4 to 8. When the Split Layer is set to 8, the stolen model achieves 100\% accuracy on clean samples and 100\% fingerprint verification success rate. In CIFAR-10, the Split Layer is set from 6 to 14. Setting the Split Layer to 14 results in 71.3\% accuracy on clean samples and 100\% fingerprint verification success rate in the stolen model. In ImageNet, the Split Layer is set from 30 to 38. Setting the Split Layer to 36 results in 83.1\% accuracy on clean samples and 100\% fingerprint verification success rate in the stolen model. Specifically, as the Split Layer is moved towards deeper layers of the model, it provides more detailed gradient information that ensures the ability to distinguish the fingerprint, hereby improving the $FVSR$ of the model. This demonstrates that the proposed fingerprint scheme remains effective in model copyright verification, with the fingerprint still being effective under label inference attacks.

\begin{figure}[!htbp]
\centering
\includegraphics[width=1\textwidth]{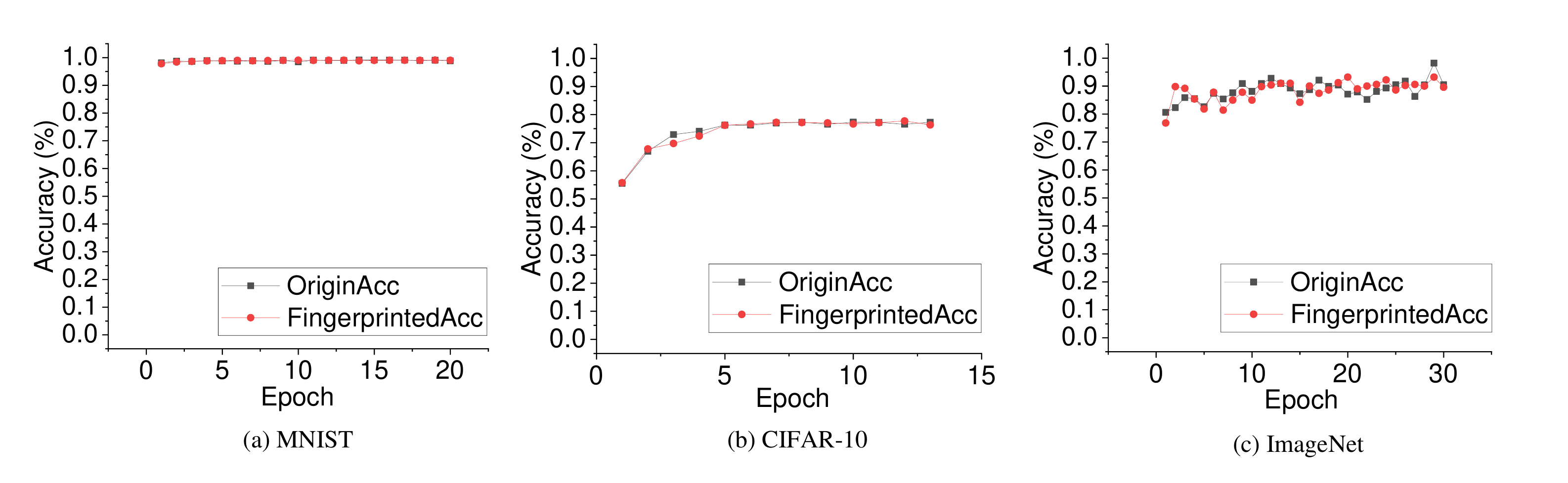}
\caption{The impact of the number of epochs on Accuracy}
\label{fig:epochs_Acc}
\end{figure}

\textbf{Epochs}: We conduct experiments to evaluate the impact of different epoch settings and obtain the following results. Figure \ref{fig:epochs_Acc} illustrates the impact of adjusting the number of epochs on the accuracy of the original and fingerprinted models. The results indicate that increasing the number of epochs enhances accuracy, leading to better overall performance. Specifically, for the MNIST and CIFAR-10 datasets, with epochs set to 16 and 10 respectively, $AccDrop$ is negligible. For the ImageNet dataset, setting epochs to 26 results in negligible $AccDrop$.

\begin{figure}[!htbp]
\centering
\includegraphics[width=1\textwidth]{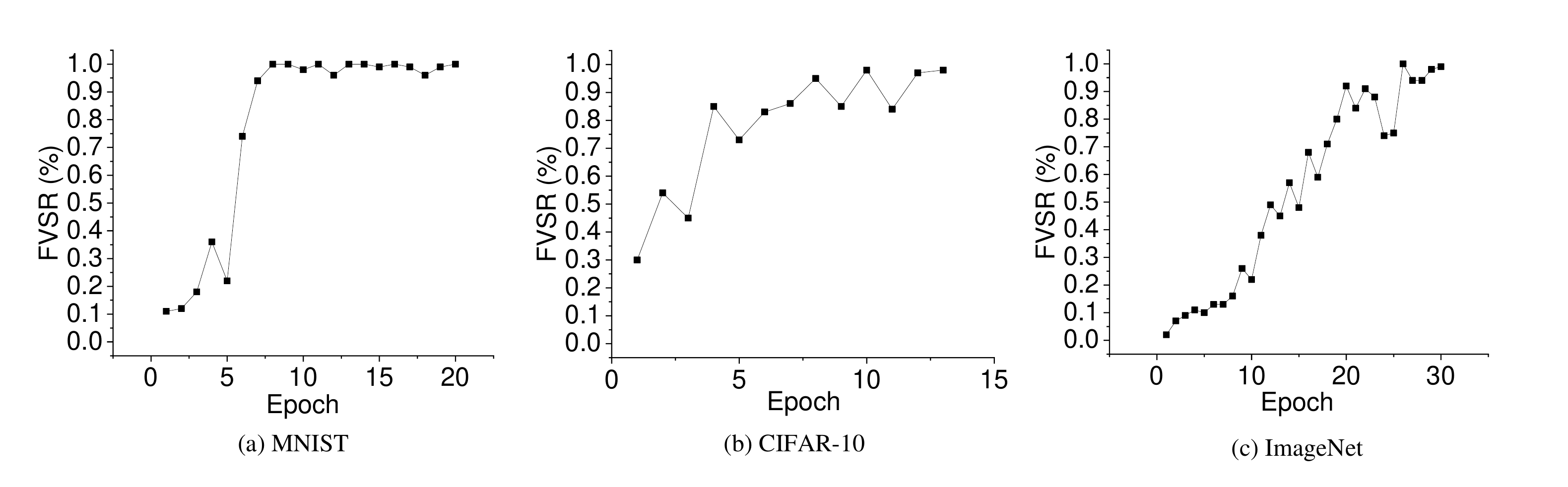}
\caption{The impact of the number of epochs on FVSR}
\label{fig:epochs_FVSR}
\end{figure}

Figure \ref{fig:epochs_FVSR} demonstrates the effect of different epoch settings on the $FVSR$. The $FVSR$ shows an overall increasing trend as the epoch increases.
For the MNIST and CIFAR-10 datasets, with epochs set to 16 and 10 respectively, the fingerprint verification success rate is nearly 100\%. Similarly, for the ImageNet dataset, setting epochs to 26 results in a 100\% fingerprint verification success rate.

\begin{figure}[!htbp]
\centering
\includegraphics[width=1\textwidth]{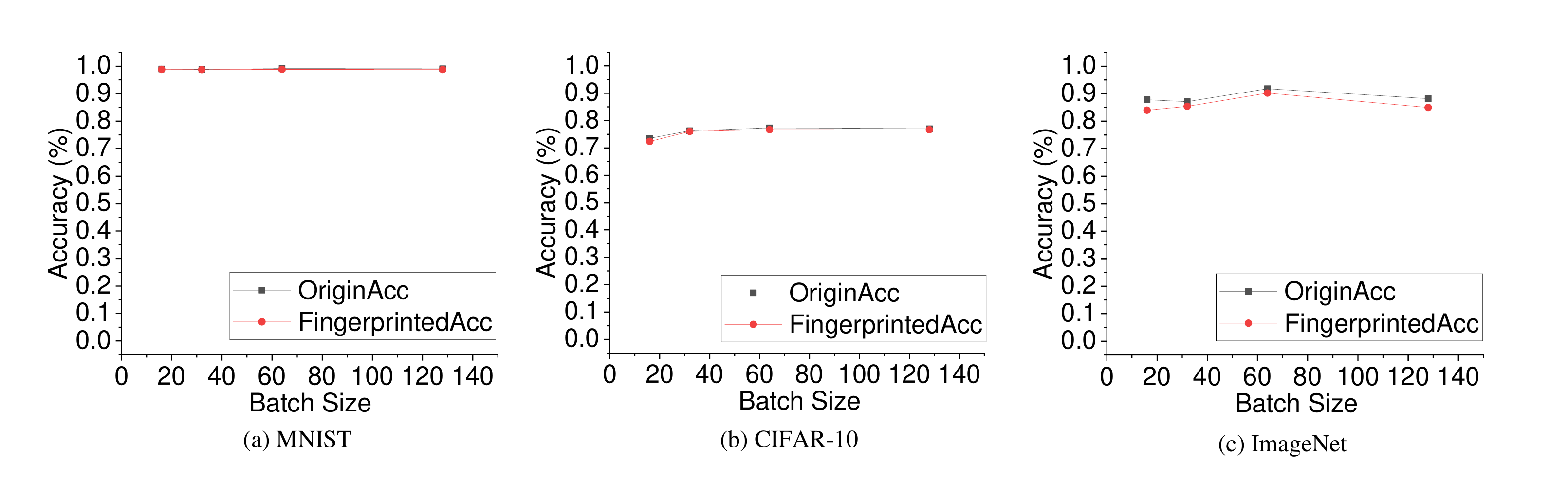}
\caption{The impact of the number of Batch Sizes on Accuracy}
\label{fig:batchsize_Acc}
\end{figure}

\textbf{Batch Size}: Batch size is a common hyperparameter in deep learning that defines the number of samples processed in each iteration of training. Therefore, we also examine the effect of different batch sizes. Specifically, we selected four batch size configurations for experimentation: 16, 32, 64, and 128. The impact of different batch sizes on model accuracy is shown in Figure \ref{fig:batchsize_Acc}. The overall accuracy remains stable across different batch sizes. For the MNIST dataset, when the batch size is set to 64, the \emph{Original model} achieves 99.14\% accuracy, while the \emph{Fingerprinted model} reaches 98.83\%. For the CIFAR-10 dataset, with a batch size of 64, the \emph{Original model} achieves 77.33\% accuracy and the \emph{Fingerprinted model} achieves 76.68\%. In the ImageNet dataset, a batch size of 64 results in 91.80\% accuracy for the \emph{Original model} and 90.20\% accuracy for the \emph{Fingerprinted model}. These results indicate that increasing the batch size does not significantly impact the accuracy, as the values remain consistently high across all datasets.

\begin{figure}[!htbp]
\centering
\includegraphics[width=1\textwidth]{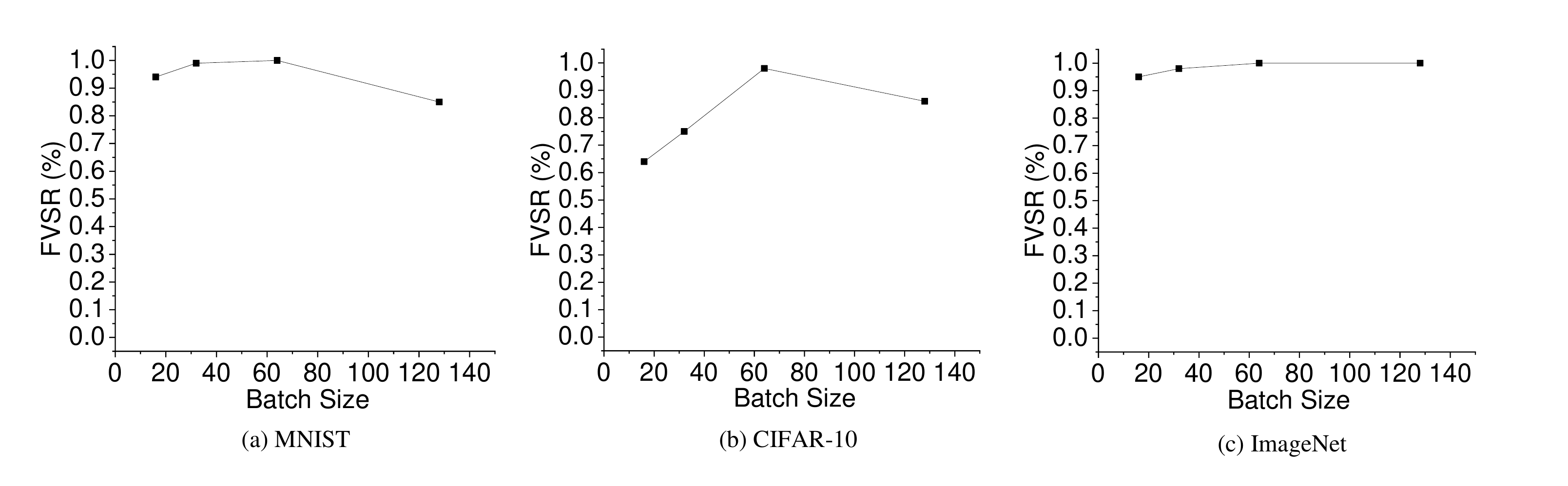}
\caption{The impact of the number of Batch Sizes on FVSR}
\label{fig:batchsize_FVSR}
\end{figure}

Figure \ref{fig:batchsize_FVSR} illustrates the impact of these batch size settings (16, 32, 64, and 128) on the $FVSR$. Changes in batch size cause minor fluctuations. For MNIST, with a batch size of 64, the fingerprint verification success rate is 100\%. For CIFAR-10, with the batch size setting to 64, the fingerprint verification success rate is 98\%. While for ImageNet, using a batch size of 64, it reaches 100\% $FVSR$. Compared to other batch size settings, the batch size of 64 yields slight improvements in $FVSR$ across all three datasets. These results indicate that selecting an appropriate batch size helps maintain a high $FVSR$ across all datasets.

\subsection{Robustness analysis}
\noindent\hspace{2em}\textbf{Label Inference Attacks \cite{Erdogan2022} Against Split Learning}: In the SplitNN structure described in Section \ref{subsec2}, label inference attacks may occur in Figures \ref{fig:f2}, \ref{fig:f3}. In these two scenarios, \emph{attacker} lacks direct access to the training data and can only intercept the gradient information transmitted between the \emph{server} and the \emph{client}. The \emph{attacker} is knowledgeable about the model architecture but does not have access to the model's parameters or the training set. The attack is carried out by intercepting the gradients exchanged between the \emph{server} and the \emph{client}, and the \emph{attacker} infers the label based on the intercepted gradients. Then he retrains a model with similar performance to the \emph{Original model}, using the inferred labels. This type of attack is known as the label inference attack in the context of Split Learning \cite{Erdogan2022}.

The fingerprint scheme proposed in this paper integrates adversarial examples as fingerprints into the training set, aiming to verify model copyright. We perform label inference attacks on the fingerprinted model, with the results shown in Figure \ref{fig:split_layer_attack}. Both the accuracy and $FVSR$ increase with the change of the position of the split layer. In the MNIST dataset, when the split layer is set to 8, the stolen model achieves 100\% accuracy on clean samples, and the $FVSR$ is also 100\%. Similarly, in the CIFAR-10 dataset, with the split layer positioned at 14, the stolen model reaches 71.30\% accuracy on clean samples, closely matching the accuracy of the fingerprinted model, and maintains 100\% $FVSR$. In the ImageNet dataset, where the split layer is set to 36, the stolen model attains 88.20\% accuracy on clean samples and 100\% $FVSR$.
These results suggest that even under label inference attacks, the fingerprint verification remains successful, indicating that the proposed fingerprint scheme is robust to such attacks and can effectively protect model copyrights in Split Learning.

\textbf{Pruning} \cite{tsuzuku2018}: We perform a pruning attack and re-evaluated the model. Table \ref{tab:pruning1} and Table \ref{tab:pruning2} present the changes in
accuracy and $FVSR$ of the fingerprinted model under pruning attack. In MNIST, at pruning rates of 10\% and 30\%, the accuracy and $FVSR$ of the fingerprinted model remain relatively high compared to the original model. At a pruning rate of 10\%, the accuracy of the fingerprinted model is 99\% and the $FVSR$ is 100\%. At a pruning rate of 30\%, the accuracy remains at 99\%, while the $FVSR$ slightly decreases to 96\%. As the pruning rate increases, the accuracy remains almost unchanged, while the $FVSR$ shows a slight decrease. Specifically, at a pruning rate of 50\%, the accuracy stays at 99\%, and $FVSR$ drops to 90\%. At a pruning rate of 70\%, the accuracy remains at 98.79\%, with $FVSR$ decreasing to 85\%. Under normal circumstances, attackers are likely to avoid excessive pruning to ensure the performance. Thus, our scheme can still maintain a high $FVSR$.

\begin{table*}[!htbp]
\centering
\caption{Accuracy and FVSR of Fingerprinted Model under varying levels of pruning attack in MNIST}
\begin{tabular}{@{}lll@{}}
\toprule
Pruning & $Accuracy$ & $FVSR$ \\
\midrule
10.00\% & 99.00\%  & 100\% \\
30.00\% & 99.00\%  & 96.00\% \\
50.00\% & 99.00\%  & 90.00\% \\
70.00\% & 98.79\% & 85.00\% \\
\botrule
\end{tabular}
\label{tab:pruning1}
\end{table*}

Table \ref{tab:pruning2} presents the impact of pruning rates on accuracy and fingerprint verification success rate in ImageNet. The results show that as the pruning rate increases, both accuracy and fingerprint verification success rate experience slight declines. With a pruning rate of 30\%, the accuracy on clean samples is 88\%, while the $FVSR$ is 98\%. The performance difference compared to the original model is negligible. At a pruning rate of 70\%, the accuracy gently decreases to 84.2\%, and the fingerprint verification success rate reduces to 91\%. Overall, as pruning rates increase, the accuracy of the model exhibits some degradation, and attackers are likely to avoid excessive pruning to maintain relatively high accuracy and preserve the overall performance of the model. This suggests that the model remains robust to pruning attacks and can still effectively recognize fingerprints.

\begin{table*}[!htbp]
\centering
\caption{Accuracy and FVSR of Fingerprinted Model under varying levels of pruning attack in ImageNet}
\begin{tabular}{@{}lll@{}}
\toprule
Pruning & $Accuracy$ & $FVSR$ \\
\midrule
10.00\% & 89.00\% & 100\% \\
30.00\% & 88.00\% & 98.00\% \\
50.00\% & 85.00\% & 94.00\% \\
70.00\% & 84.20\% & 91.00\% \\
\botrule
\end{tabular}
\label{tab:pruning2}
\end{table*}

\subsection{Comparison}
\noindent\hspace{2em}In this section, we compare our work with related studies across various machine learning frameworks, including deep learning, federated learning, and Split Learning. We focus on robustness, accuracy degradation, and $FVSR$. This paper presents a fingerprint scheme based on adversarial examples specifically designed for the Split Learning scenario, which distinguishes it from studies focused on deep learning models \cite{Lv2023,Kallas2022} or Federated Learning \cite{Tekgul2021,Li2022}.

Compared to most schemes, our scheme maintains strong robustness against label inference attack and pruning attack. Furthermore, our scheme still maintains high performance, with only a 0.31\% reduction in accuracy after embedding
fingerprints. This subtle performance degradation is similar in other work \cite{Kallas2022} in Deep Learning or works \cite{Tekgul2021, Li2022} in Federated Learning. Our work demonstrates 100\% $FVSR$. This is on par with other approaches, such as \cite{Li2022}, which also achieves 100\% $FVSR$ in federated learning. However, it outperforms works like \cite{Kallas2022}, where the $FVSR$ is 92.50\%. Therefore, our approach ensures robust copyright protection with strong resistance to label inference attacks, while maintaining low complexity and high accuracy, making it well-suited for Split Learning.
\begin{table*}[!ht]
\caption{Comparison with existing works}
\resizebox{\textwidth}{!}{
\begin{tabular}{@{}lllll@{}}
\toprule
Works & Scenario & Robustness & $AccDrop$ & $FVSR$ \\
\midrule
\cite{Lv2023} & Deep Learning & Pruning, Fine-tuning & 0.14\% & 99.9\% \\
\cite{Kallas2022} & Deep Learning & Pruning, Fine-tuning, JPEG attack & 0.24\% & 92.50\% \\
\cite{Tekgul2021} & Federated learning & Pruning, Fine-tuning, reverse engineering & 0.17\% & 99\% \\
\cite{Li2022} & Federated learning & Pruning, Fine-tuning & 0.26\% & 100\% \\
Ours & Split Learning & Pruning, Label inference attack & 0.31\% & 100\% \\
\botrule
\end{tabular}}
\label{tab:comparison}
\end{table*}

\section{Conclusion}\label{conclusion}
\noindent\hspace{2em}In this paper, we propose the first copyright protection scheme for Split Learning models, which embeds adversarial examples as fingerprints into the model's training data. During training process, the fingerprint is embedded in the model for copyright verification. Our experiments demonstrate the effectiveness of the proposed fingerprint scheme and confirm its robustness against attacks. Future research will focus on active copyright protection mechanisms for Split Learning model.

\section*{Acknowledgment}
This work is supported by the National Natural Science Foundation of China (No. 62372231), Aeronautical Science Foundation (No. 2022Z071052008) and CCF-NSFOCUS Kun-Peng Scientific Research Fund (No. CCF-NSFOCUS 2024002).

\section*{Declarations}
\bmhead{Data availability}
The data is available upon reasonable request.

\bibliography{ref}

\end{document}